\documentclass[conference]{IEEEtran}
\usepackage{cite}
\usepackage{amsmath,amssymb,amsfonts}
\usepackage{algorithmic}
\usepackage{graphicx}
\usepackage{textcomp}
\usepackage{xcolor}
\usepackage{booktabs}
\usepackage{xurl} %fix long URLS in references
\usepackage{paralist} %inline summarization
\usepackage{caption}
\usepackage{dirtytalk}

\newcommand{\opret}{OP\_RETURN}

\begin{document}

\title{Bitcoin Battle:\\ Burning Bitcoin for Geopolitical Fun and Profit}
%Bitcoin Black Ops
%Bitcoin Blowback

\author{
    \IEEEauthorblockN{Kris Oosthoek}
    \IEEEauthorblockA{
        \textit{Delft University of Technology}\\Delft, The Netherlands \\
        k.oosthoek@tudelft.nl}
    \and
    \IEEEauthorblockN{Kelvin Lubbertsen}
    \IEEEauthorblockA{
        \textit{Delft University of Technology}\\Delft, The Netherlands \\
        k.j.m.lubbertsen@tudelft.nl}
    \and
    \IEEEauthorblockN{Georgios Smaragdakis}
    \IEEEauthorblockA{
        \textit{Delft University of Technology}\\Delft, The Netherlands \\
        g.smaragdakis@tudelft.nl}
}

\maketitle

\begin{abstract}
This study empirically analyzes the transaction activity of Bitcoin addresses linked to Russian intelligence services, which have liquidated over 7 Bitcoin (BTC), i.e., equivalent to approximately US\$300,000 based on the exchange rate at the time. Our investigation begins with an observed anomaly in transaction outputs featuring the Bitcoin Script \opret\ operation code, tied to input addresses identified by cyber threat intelligence sources and court documents as belonging to Russian intelligence agencies. We explore how an unauthorized entity appears to have gained control of the associated private keys, with messages embedded in the \opret\ outputs confirming the seizure. Tracing the funds’ origins, we connect them to cryptocurrency mixers and establish a link to the Russian ransomware group Conti, implicating intelligence service involvement. This analysis represents one of the first empirical studies of large-scale Bitcoin misuse by nation-state cyber actors.
\end{abstract}

%Storyline:
%- OP\_RETURN anomaly
%- OP\_RETURN messages
%- messages
%- linking messages to BTC addresses
%- open source support for linking these addresses to actors
%- what can we see in the days before the 'attack'? (in Feb and March?) any prep TXs on these addresses?
%- other activity from these addresses (ransomware? hacking?) - look into Crystal
% was there any money left? or was this a way to rob the agencies of all their funds?
% which could also be done by just seizing the money, so this seems to be a more ethic actor


\section{Introduction}

Bitcoin has emerged as a significant instrument of state power in cyber-conflicts, with state actors and their affiliates leveraging the cryptocurrency for strategic advantage. Multiple sources confirm that governments and associated entities exploit Bitcoin’s decentralized nature to circumvent traditional financial systems. For instance, media reports indicate that Iran's Islamic Revolutionary Guard Corps (IRGC) has engaged in Bitcoin mining to mitigate the impact of international sanctions~\cite{terrorist-crypto} and has utilized Bitcoin transactions to fund proxy groups in the Middle East, such as Hamas and Hezbollah~\cite{hamas-crypto}. North Korea has been involved in numerous hacks of cryptocurrency exchanges, amassing approximately US\$3 billion over a six-year span~\cite{nk-hacking}. In 2024, Russia legalized crypto mining~\cite{mining-cnbc}, employing it as a tool to sustain international trade amid economic sanctions~\cite{russia-foreign-trade}, and Russian ransomware groups have been linked to co-opting with intelligence services. The Conti group, which has generated US\$300 million in ransomware profits, is reportedly connected to the Federal Security Service of the Russian Federation (FSB)~\cite{conti-fsb}. Likewise, members of Evil Corp, associated with tge LockBit and BitPaymer ransomware strains, have FSB links either through prior employment~\cite{evilcorp-dridex} or familial connections~\cite{evilcorp-family}.

In addition to its use by adversarial states, Bitcoin has drawn the attention of the United States as a target for surveillance. Leaked documents reveal that, as early as 2013, the U.S. employed signals intelligence to track Bitcoin senders and receivers~\cite{btc-sigint}. Conversely, Russia's use of Bitcoin in 
cyber-espionage campaigns is also well-documented. During the 2016 hack of the Democratic National Committee, attackers linked to Russian intelligence leveraged Bitcoin to purchase servers and domains~\cite{dnc-complaint}. Similarly, in the SolarWinds supply-chain attack in 2020, alleged Russian state-sponsored actors used Bitcoin to purchase infrastructure~\cite{nsa-solarwinds}. In both instances, Bitcoin’s pseudonymity enabled these actors to obscure their operations, hindering law enforcement efforts to trace activities through traditional financial channels. However, academic work on nation-state actors' use of cryptocurrency is limited compared to cybercriminal Bitcoin usage.

Beyond analyzing financial transactions, which has become commonplace through blockchain analysis, investigating other blockchain events can yield interesting results. One of these is \textit{burning}, which can manifest itself in three ways. First, Bitcoin (BTC) is forever lost, that is, burned, if a miner fails to claim the block reward in the \textit{coinbase} transaction. Second, BTC is removed from circulation when sent to a non-existing address as it distorts the cryptographic rule in which each address is a public key controlled with a secret key. Third, a non-existing or fabricated address is essentially a public key for which no private key exists, making the BTC unspendable.

Our analysis focuses on the third burn option named \opret\, an operation code (opcode) within the Bitcoin blockchain script to mark a transaction output as invalid. Also called \textit{nulldata} field, it can add up to 80 bytes of arbitrary data to the transaction stored permanently on the blockchain. Sending to \opret\ creates an unspendable output, removing funds from circulation~\cite{btcwiki-opret}. The \opret\ opcode was added to Bitcoin in March 2014 as part of the Bitcoin Core version 0.9.0 release~\cite{bitcoin090}. It was originally introduced to provide a way for developers to store small amounts of data on the blockchain to minimize network impact and avoid known downsides of previous data storage methods in Bitcoin transactions. With its introduction, Bitcoin Core developers did not endorse storing data on the blockchain, as it would bloat the \textit{unspent transaction output} (UTXO) database. Speculatively, this is why the feature was designed to render any amount of BTC unspendable when used. 

Technically, an \opret\ output with a zero value can be included in a larger transaction with multiple outputs. There is no requirement to attach a meaningful amount of \textit{satoshis} (one one-hundred-millionth of a BTC, the smallest unit of account) to the \opret\ output, but it is used rarely; the mean value of BTC sent daily to \opret\ from January 1, 2023, until August 1, 2024, was 0.00805 BTC.

\opret\ allows users to store small amounts of data on the Bitcoin blockchain, taking advantage of the irreversible and immutable nature of the blockchain. Once the data are included in a block, it becomes a permanent tamper-resistant record that is accessible to anyone inspecting the blockchain. As an example, El Salvador's announcement of accepting Bitcoin \cite{elsalvador} as legal tender was recorded on the blockchain using \opret.\footnote{Transaction hash: {\tt cb01ea705494ce66d7e5b7cb51bb5b39b8e8ce\\31e168d1bd7dda253af359cc77}} 
The introduction of Taproot in 2021 significantly improved Bitcoin's scripting capabilities, allowing more efficient use of transaction space and greater data storage flexibility \cite{taproot}. This inadvertently facilitated the creation of the Ordinals protocol, which uses Taproot's expanded data storage features to inscribe arbitrary data, such as text or images, onto individual satoshis, without relying on \opret.

Hence accounts of usage of \opret\ instructions are scarce, let alone coverage of their use in the context of nation-state actors' cyber operations. The only coverage we are aware of is by Sophos and Google, which in 2020~\cite{glupteba-sophos} and 2022~\cite{glupteba-tag} reported on malware being controlled through \opret\ outputs. In addition to analysis by cybersecurity firms, this is the first academic account of nation-state actors' usage of \opret\ outputs in order to indefinitely burn a record amount of Bitcoin.

\vspace{1em}
\noindent The contributions of this paper are as follows:
\begin{itemize}
    \item We analyze the largest event of rendering BTC unspendable in the history of Bitcoin.
    \item We characterize the events in a geopolitical and cyber context, linking them to cyber espionage.
    \item We link wallets to cyber espionage, ransomware and hacking based on open and semi-open sources.
    \item We analyze significant simultaneous activity in over 60,000 automated fractional payment transactions between these wallets.
    \item We release 1,011 labeled wallet addresses to the community based on the analysis of this work \cite{btc-battle-github}.
\end{itemize}

\vspace{1em}
The remainder of this paper is structured as follows. Section 2 provides an overview of the related work on the \opret\ instruction and exploits thereof. Section 3 provides an overview of Russian Cyber Operations. Section 4 describes the methodology of our analysis, including our initial findings. Section 5 attributes the wallets observed in the dataset to Russian intelligence services based on open sources. Section 6 explores simultaneous fractional payment activity. Section 7 summarizes our findings.

%Free Use: You can't use OP_RETURN "for free" in the sense of bypassing fees or sending BTC for free. You still have to pay transaction fees to miners for including the transaction in a block, and any BTC tied to the OP_RETURN output is burned. However, there is no requirement to attach any meaningful amount of BTC to the OP_RETURN output (it can be set to 0 BTC), so the primary cost is the transaction fee rather than the amount of BTC being "burned."
\section{Related Work}

For the related work for our analysis, we consider 
academic contributions examining the \opret\ instruction, its associated exploits, and pertinent discussions in popular Bitcoin blogs and media.
%three categories of related work: 

Bartoletti and Pompianu~\cite{bartoletti} conducted a comprehensive study of the types and volumes of metadata embedded in \opret\ outputs, identifying their use in applications such as timestamping, asset tracking, and data anchoring. Other research has focused specifically on exploitation of the opcode for illicit purposes. Böck et al.~\cite{opret-ecrime} reported on blockchain-based botnets, where botmasters leverage blockchains' decentralized and censorship-resistant nature to establish command and control (C\&C) channels. The authors assessed that the adversary's benefit of using blockchains for C\&C is primarily resistance to law enforcement take-down but that its adoption is tempered by financial costs and technical limitations. Similarly, Matzutt et al.~\cite{matzutt} examined how the metadata in Bitcoin is used to store potentially harmful or illegal content. Their analysis uncovered more than 1,600 embedded files, some containing objectionable material such as links to illegal content, which could make possession of the blockchain illegal in certain jurisdictions. Lastly, Narula and Narula~\cite{deadbolt} analyzed the Deadbolt ransomware, which, upon receiving payment, releases decryption keys on the blockchain through \opret\ transactions.

Coverage of \opret\ is scarce in popular media. The \opret\ lemma of the official Bitcoin Wiki is relatively short \cite{btcwiki-opret}. As discussed earlier, the coverage of the Glupteba malware by Sophos in 2020~\cite{glupteba-sophos} and Google's Threat Analysis Group in 2022~\cite{glupteba-tag} described its use of \opret\ outputs. Both described Glupteba as a backdoor capable of stealing sensitive information, mining cryptocurrency, and enrolling infected devices in a botnet. The malware's operators embed encrypted data within \opret\ outputs, pointing to new C\&C servers. This allows the malware to recover quickly even if one set of C\&C servers is compromised or taken offline. The malware continuously monitors the blockchain for new transactions containing specific \opret\ data, ensuring that it can update its C\&C addresses dynamically and autonomously. By not relying on a fixed domain name or a centralized server for C\&C, the malware authors mitigate the risk of traditional C\&C infrastructure being blocked by defenders or taken down by law enforcement.

Certain elements of our analysis concerning the evaporation (or \say{burning}) of BTC align with observations in a blog post published in April 2023 by blockchain analysis firm Chainalysis~\cite{chainalysis-blog}. The post offers a visual overview of transaction activity and references the \opret\ messages. However it provides only a cursory examination of the associated campaign. In contrast, this paper conducts a detailed investigation into BTC burning by nation-state cyber actors. Drawing on a diverse array of sources, we correlate these activities with geopolitical events over time, providing a comprehensive analysis of their strategic significance.
\section{Russian Cyber Operations}
%this needs references
%https://en.wikipedia.org/wiki/Cyberwarfare_by_Russia

Our research investigates a specific class of Bitcoin transaction metadata, with particular emphasis on references to the GRU, FSB and SVR. This section explains their role in cyber operations, not self-evident to the blockchain and cryptocurrency community. The SVR (Foreign Intelligence Service), FSB (Federal Security Service), and GRU (Main Intelligence Directorate) are Russian Federation government's intelligence agencies that have become focal points in cyber security, espionage, and cyber warfare in a broader sense.
%\vspace{1em}

\subsection{SVR}

The \textit{SVR} focuses on long-term intelligence collection and offensive operations. Its targets include diplomatic organizations, technology companies, international organizations, and defense contractors \cite{nsa-svr}. A cyber-espionage campaign attributed to the SVR is the attack on the software company \textit{SolarWinds} in 2020, where an SVR cyber group, \textit{APT29}, compromised SolarWind's software. The attack impacted several U.S. government agencies and tech companies using SolarWind's software. The breach allowed the SVR to monitor internal communications and exfiltrate sensitive information for months before being detected \cite{nsa-solarwinds}.

The ongoing operations of SVR-attributed actors APT29, Midnight Blizzard, Dukes, and Cozy Bear have a wider range of targets. With evolving TTPs, the actor has been observed to transform the operation from compromising on-premises networks to cloud infrastructure \cite{gchq-svr}. SVR operations focus on stealth and persistence, with long-term intelligence gain techniques aimed at political and economic advantages.
%\vspace{1em}

\subsection{FSB}

The \textit{FSB} is primarily responsible for domestic security and counterintelligence, which extends to the cyber domain. In 2016, FSB-linked hackers launched \textit{Armageddon}, a long-running cyber-espionage operation targeting Ukraine \cite{armageddon}. The operation carried out amidst the ongoing conflict between Russia and Ukraine, involved spear-phishing attacks and malware designed to steal military and government secrets.

In 2017, the \textit{NotPetya} attack, attributed to the FSB and GRU, targeted Ukraine but caused global disruption \cite{notpetya}. NotPetya exploited a leaked backdoor developed by the US National Security Agency (NSA), causing damage to companies including Maersk, FedEx, and Merck. Disguised as ransomware, it was intended to erase data and disrupt operations, causing an estimated US\$10 billion in damages worldwide.

Two \textit{FSB} agents have been charged by the FBI for conducting a data breach of Yahoo! in 2016, in which more than 500 million Yahoo! accounts were compromised~\cite{yahoo}. The attack, in collaboration with cybercriminals, exposed user data and demonstrated the FSB's reliance on criminal proxies to carry out large-scale cyber operations. More recently, the National Crime Agency (NCA) has described how the FSB has been co-opting {\it Evil Corp}, a cybercrime group for its own malicious cyber activity~\cite{evilcorp}. The FSB tasked Evil Corp with conducting cyber attacks and
espionage operations against NATO allies.
%\vspace{1em}

\subsection{GRU}

The \textit{GRU} has been implicated in disruptive and aggressive cyber attacks. In the lead-up to the 2016 US presidential election, GRU cyber units \textit{Unit 26165} and \textit{Unit 74455} orchestrated a series of hacks targeting the Democratic National Committee (DNC) and the Democratic Congressional Campaign Committee (DCCC) \cite{dnc-complaint}. Emails stolen during these hacks were later leaked through WikiLeaks, allegedly with the intention of influencing the US election outcome.

In the same year, the GRU conducted \textit{Ghostwriter}, a disinformation operation targeting Eastern European countries. It involved hacking news outlets and altering articles to spread pro-Russian narratives, particularly in countries like Lithuania, Latvia, and Poland \cite{ghostwriter}. The GRU has also been held responsible for a cyber operation called \textit{Olympic Destroyer}, targeting the 2018 Winter Olympics in South Korea. The attackers intended to disrupt the IT systems that support the event, including Wi-Fi networks and ticketing services \cite{olympic}. Although the attack was designed to seem like it originated in North Korea, forensic analysis linked the operation to GRU's Unit 74455 (Sandworm), which targets critical infrastructure.

In 2015 and 2016, the GRU breached Ukraine's power grid, causing blackouts in several regions. The \textit{Industroyer malware} (or \textit{CrashOverride}) was designed to disrupt industrial control systems (ICS) used in power grids \cite{crashoverride}. Together with Stuxnet~\cite{stuxnet}, this is one of the few instances in which a cyber operation caused physical disruption of critical infrastructure.
\section{An OP\_RETURN Anomaly}

%\vspace{0.2cm}
\begin{table}[t] %[htbp!]
\centering
\begin{tabular}{cc}
\toprule
\textbf{Date} & \textbf{OP\_RETURN Value} \\
\midrule
\textit{2022-02-18} & \textit{4.167737} \\
\textit{2022-02-12} & \textit{3.691136} \\
2021-02-14 & 3.548904 \\
2021-02-13 & 2.591002 \\
2020-10-11 & 2.405650 \\
2021-02-18 & 2.345250 \\
2020-10-23 & 1.988000 \\
2021-02-15 & 1.873764 \\
2020-10-12 & 1.746181 \\
2015-09-11 & 1.565100 \\
\bottomrule
\end{tabular}
%\vspace{0.5cm}
\caption{Top Days with BTC expenditures to OP\_RETURN \\ (dates relevant to this analysis in italic).}
\label{opret-daily}
\end{table}

To facilitate an exploratory investigation of transactions containing \opret\ opcodes (from here called OP\_RETURN), we parsed all historical transactions from a Bitcoin full node up until August 1, 2024. Bitcoin transactions can be appended with small script operations. Bitcoin Script is a stack-based programming language that defines the conditions under which Bitcoin transactions can be spent, enabling features such as multi-signature and time-locks through a set of operation codes (opcodes), instructing the blockchain what to do. If the script starts with the hexadecimal equivalent of \opret\ (i.e., {\tt 6a}), the opcode for \opret\, is identified as an \opret\ output. The use of \opret\ does not necessarily burn the bitcoin transacted. It is possible to assign a zero value to the output. Any value other than null indicates that the corresponding amount of satoshi (one hundred millionth of a Bitcoin) will be burned.

\subsection{Methodology}
To gather the \opret\ binary metadata central to this analysis, we used a full node running Bitcoin Core version 26.2, synchronized with the blockchain until July 2024. To facilitate a quick analysis of transactions containing \opret\ opcodes (from here called OP\_RETURN), we parsed all historical transactions until August 1, 2024. We used the \textit{blockchain-parser} library \cite{blockparser} to convert Bitcoin-native raw blk*****.dat files, based on the {\tt getrawtransaction} RPC command to plain text. The resulting text files were parsed on the fly for transactions containing \opret\ opcodes using regular expressions, the output of which was pushed to a DataFrame object for efficient query and data manipulation, run in memory on a workstation with significant (128 GB) RAM.

Upon obtaining the initial set of suspicious transaction hashes, the parsed Bitcoin data was crawled to discover associated inputs, outputs, and additional transactions. Data was obtained from the free Blockchain.com API~\cite{blockchain} for a sample of transaction hashes. This allowed for the verification of the results from our parser, serving as a double check. We were provided access to, and we used the Graphsense API maintained by Iknaio for its address clustering capabilities, based on the co-spend heuristic~\cite{meiklejohn2013fistful} to discover joint address ownership. We were also provided access to address labels by Scorechain~\cite{scorechain} under an academic license to identify links with counterparties.
A full list of the addresses in our dataset and their derived labels (GRU, SVR, FSB) is available online on GitHub and IPFS (to be made public with the publication of the paper).

\begin{figure}[!tpb]
\centerline{\includegraphics[width=0.5\textwidth]{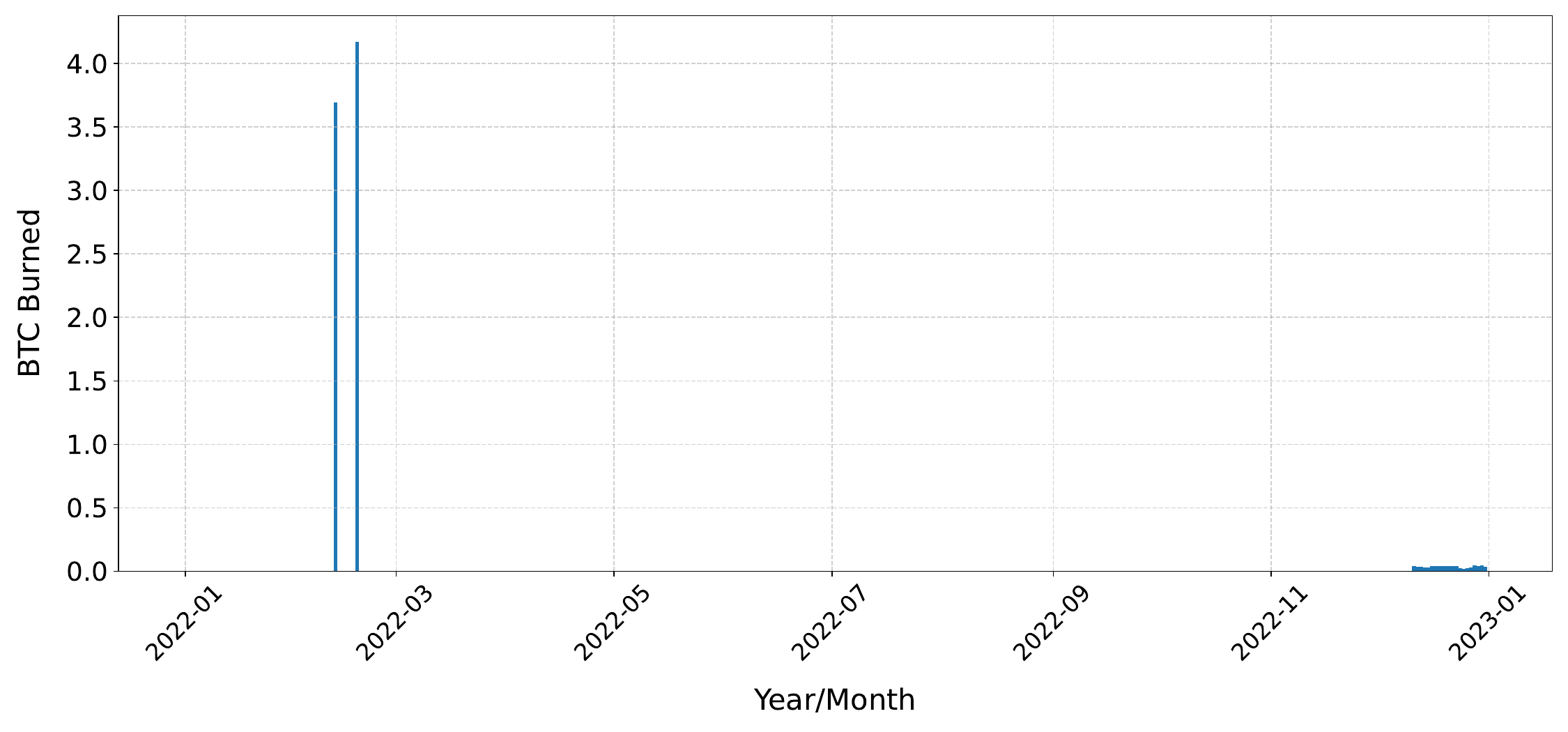}}
\caption{Fig. 1: Time series of BTC Burned on OP\_RETURN in 2022.}
\label{opret-2022}
\end{figure}

\begin{figure*}[htbp]
\centering
\captionsetup[figure]{labelformat=empty}
\caption{TABLE II: Summary of Transactions containing OP\_RETURN Outputs}
%\vspace{-0.75cm}
%table included in separate pdf to avoid formatting issues due to Cyrillic text
% trim: left bottom right top
\includegraphics[trim=0.3cm 14cm 0.4cm 13cm, width=\textwidth]{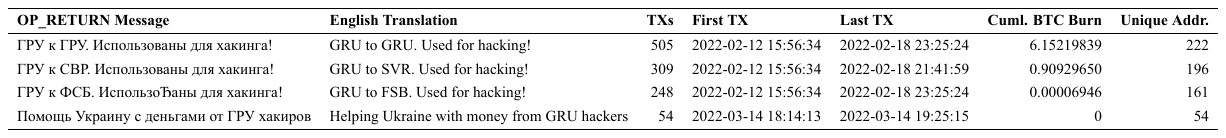}
\label{callouts}
\end{figure*}

With a feature that potentially renders funds irrecoverable, the use of \opret\ is relatively limited. The mean value of BTC sent daily with \opret\ from January 1, 2023, to August 1, 2024, is 0.00805 BTC. Only on a few historical dates does the total value peak above a single Bitcoin. Table \ref{opret-daily} provides an overview of the historical daily top spending with OP\_RETURN, based on aggregated daily non-zero outputs containing the \opret\ opcode. This overview made us decide to dig deeper into the two top dates, as these were also the most recent.

We found that on February 12, 2022, i.e., twelve days before Russia's invasion of Ukraine (on February 24, 2022) after months of military preparations, a total of 583 \opret\ transactions took place. A couple of days later, on February 18, another batch of 210 transactions followed. All transactions had either of three \opret\ messages attached. The fact that the \opret\ campaign took place less than two weeks before the actual invasion should not be regarded in isolation. We argue that it rather must be understood as part of a broader socio-political and technological context that includes the intersection of cryptocurrency adoption and hacking by state actors and geopolitical instability.

Figure~\ref{opret-2022} provides a visual impression of the significance of the amount of BTC burned on these two dates, relative to \opret\ disbursement during the rest of the year 2022.
We queried the output of our custom parser to discover additional transactions from the addresses involved in the initial set of transactions that took place on February 12 and 18. Based on that, we identified 11 additional transactions on March 14th (i.e., after the Russian invasion of Ukraine) with a different \opret\ message, sending funds to the official Bitcoin donation address of the Ukrainian Armed Forces \cite{ukraine}. However, the \opret\ outputs on this date had 0 BTC attached, as shown in Table II.

As shown in Table II, %hardcoded
the transactions included \opret\ instructions that with hexadecimal values representing Cyrillic characters, which can be translated into four distinct callouts:  \begin{inparaenum}[(i)]
    \item GRU to SVR,
    \item GRU to FSB,
    \item GRU to GRU, and
    \item Helping Ukraine with money from GRU hackers.
\end{inparaenum}

Table II provides an overview of the Cyrillic and English versions of the messages and their count of appearances within blockchain transactions. Furthermore, the table shows timestamps for the first and last transactions per message, the cumulative amount of (fractional) Bitcoin burned by the different messages, and the number of unique addresses that participated in sending each message. Not all identified wallet addresses have engaged in burning Bitcoin. Our data set consists of 986 addresses that were either inputs or outputs in at least one \opret\ transaction. Of these 986 addresses, 275 were used as input addresses in the \opret\ transactions, burning 7.06 BTC in total. However, as will be discussed in the next section, all of the 986 addresses in our dataset engaged in sending and receiving small payment transactions.

To check the attribution of each address, we took the first sentence of each message, splitting it based on the word \say{to}. As an example, for \say{GRU to SVR}, we assumed the inputs are GRU wallets and the outputs SVR wallets. We only considered outputs, as observed in Table 2, the inputs are always GRU. For example, in transactions labeled \textit{GRU to FSB}, the supposed GRU address will appear both as input and output due to it being a change address. Judging from the messages summarized in Table II, only addresses attributed to the GRU were used as transaction output, and the SVR and FSB were used as outputs. Although, based on the outputs, SVR and FSB addresses do indeed appear in the transactions, these rather engage in payment activity during the campaign, but not \opret\ outputs.

As shown in Table II and further discussed in the next section, the 14 March, 2022 \opret\ outputs referring to helping Ukraine did not burn any Bitcoin but were accompanied by the transfer of money to Ukraine's donation address~\cite{ukraine}. For this reason, they are not represented in Table II. The difference in the total burned per date in Figure~\ref{opret-2022} and Table II can be explained by the \opret\ transactions that took place on these dates but were not associated with this campaign.

\section{Address Ownership}
In order to empirically assess ownership of the addresses, we have been looking for evidence of this in reliable sources, open to the public. Hence, we queried for sources reporting on usage of Bitcoin in Russian cyber operations. Of the addresses in the dataset, three have been publicly attributed by reliable sources to Russian intelligence agencies. This section considers these findings.

\subsection{Democratic National Committee Breach}
According to various sources, the Democratic National Committee (DNC) was hacked by Russian actors in 2018. The official indictment by a US court assesses that the actors have used Bitcoin to purchase VPN accounts, server infrastructure, and domain names \cite{indictment}. Specifically, the indictment mentioned that newly mined Bitcoin were used to fund the attack infrastructure. This is consistent with the hypothesis that the Kremlin uses the fruits of Bitcoin mining for subversion \cite{rusi-gru-bitcoin}. Although the indictment does not mention any Bitcoin address, an industry media blog post includes the address {\tt 18N9jzCDsV9ekiLW8jJSA1rXDXw1Yx4hDh} \cite{chainalysis-blog}.

The DNC hack is widely attributed to Fancy Bear and Cozy Bear \cite{dnc-attr-1, dnc-attr-2}. While Fancy Bear is linked to GRU Unit 26165, Cozy Bear is linked to the SVR \cite{indictment}. In the \opret\ messages, the aforementioned public wallet address is linked to the GRU.

\begin{figure*}
    \caption{TABLE III: General statistics of clusters}
    \centering
    \begin{tabular}{lrrrrrr}
    \toprule
        \textbf{Entity} & \textbf{\# addresses} & \textbf{\# clusters} & \textbf{Cluster size (avg)} & \textbf{Cluster size (std)} & \textbf{\# transactions (avg)} & \textbf{\# transactions (std)} \\
    \midrule
    SVR & 3 & 3 & 1.000000 & 0.000000 & 1.666667 & 0.471405 \\
    GRU & 15,856 & 872 & 16.196118 & 57.489070 & 13.371560 & 31.313955 \\
    FSB & 13 & 4 & 3.250000 & 2.277608 & 204.250000 & 2.277608 \\
    \bottomrule
    \end{tabular}
    \label{tab:cluster-statistics}
\end{figure*}

\subsection{SolarWinds Breach}
We found an archived blog by cyber incident response company HYAS, reporting on its forensic investigation of the SolarWinds hack in 2020, which mentioned two hashes of Bitcoin transactions to procure attack infrastructure \cite{hyas}. On inspection of the transaction data, we found the source addresses {\tt 1DLA46sXYps3PdS3HpGfdt9MbQpo6FytPm} and {\tt 1L5QKvh2Fc86j947rZt12rX1EFrCGb2uPf} also occurred in our dataset. We labeled these addresses as SolarWinds for further analysis.

The SolarWinds breach is publicly linked to the SVR, specifically to a group known as APT-29 (Advanced Persistent Threat 29), also referred to as Cozy Bear \cite{solarwinds-attr}. The \opret\ callouts in our dataset do also link the two wallet addresses to the SVR.

Furthermore, according to a report by Western intelligence agencies, Unit 29155 of the GRU, the 161st Specialist Training Center, has employed the WhisperGate malware against Ukranian and other NATO targets \cite{whispergate}. According to the report, the actors used Discord for the distribution and control of malware hidden as ransomware. The fake ransom note displayed by the malware listed the Bitcoin address {\tt 1AVNM68gj6PGPFcJuftKATa4WLnzg8fpfv}. Although this address does not appear in our cluster of GRU addresses, we mention it here as this  Bitcoin link to GRU cyber operations.

The transaction activity of the DNC and SolarWinds addresses prior to the callouts is typical of a sophisticated cyber actor. Exactly as reported for ransomware syndicates \cite{Oosthoek2023}, the addresses were only used once, i.e., a deposit, followed by a payment for infrastructure. They only become active again during the \opret\ campaign reported in this analysis.

Inherent to Bitcoin's asymmetric cryptography, the \opret\ transactions must have been initiated by an actor in possession of the private keys. The one-way hash function used to generate the public-private key pair cannot be reversed. As an example, this means that the private key of the addresses implicated in the DNC and SolarWinds hacks, used to purchase the attack infrastructure, was also used for \opret\ transactions. This can be compared to a password being obtained from a password manager and then being used to act as if one is the legitimate owner.

\subsection{Address Characteristics}
As shown in Table III,
%\ref{tab:cluster-statistics}, 
we found only three addresses in our dataset labeled as belonging to SVR. As discussed earlier in this section, two of these have been publicly linked to the attack on SolarWinds by the SVR. In addition, only four addresses were labeled as FSB. This means that most addresses belong to the GRU, at least according to the callouts. Furthermore, six addresses belonging to GRU start with \textit{bc1} and thus are Bech32/SegWit addresses. All other addresses in the dataset start with `1' or `3' and thus are considered legacy addresses. Going with the hypothesis that the private keys were seized, using legacy addresses might suggest something about the software wallet type used to store the private keys.

\subsection{Label-Based Address Clustering}
With address ownership confirmed as far as possible, we applied clustering to the addresses using the co-spending heuristic first described by \cite{meiklejohn2013fistful}. This heuristic merges all input addresses in an outgoing transaction to the same entity under the assumption that all inputs have to be signed. We argue that even though a third party allegedly compromised the private key, this still gives it access to only the funds of one group, GRU, as the descriptions in the \opret\ metadata only mention GRU as the sender.

All transactions, $984$ addresses, can be labeled as associated with the GRU, SVR, or FSB. When applying co-spending, we learn that these fall into $879$ clusters (see Figure \ref{tab:cluster-statistics}). By clustering, we associate new addresses with those in the initial dataset. We learn that none of the clusters overlap between agencies; for instance, no addresses related to the GRU are also used by the FSB.

In Table III, 
%\ref{tab:cluster-statistics}, 
we describe the statistics for the cluster. We must note that to be able to observe nation-state actor's activity, we must delete transactions done by wallet hackers. We assume they have included an \opret\ output in every transaction since their motive appears to have been doxxing. That way, we can analyze transactions associated with the original owner of the wallets and, for the first time, analyze the behavior of these nation-state actors. With three addresses being linked to Russian cyber actors by official sources, which appear in the co-spend address clusters in Table III, %\ref{tab:cluster-statistics}, 
we can confidently establish that Russian actors indeed controlled these at one point.

\section{Financial Analysis}
%this focuses not on the burned BTC, but rather other financial activity where money is sent, not burned
%notebooks: primary and secondary
We distinguish between financial transactions that took place as part of the OP\_RETURN campaign, but that did not burn money. We call these payment transactions. As the actor behind the campaign also put some externally sourced funds into it, we will focus on these transaction first, which we will call funding transactions.

\subsection{Funding Transactions}
%https://chain.so/tx/BTC/96e8c84dfa9dcbe5b161c345877381f2c2e83a464c1db1e149c9b0071da9ced8
On February 1st, 2022, the likely actor behind the campaign put 1 BTC into a wallet associated with Cryptomixer.io, a well-known centralized cryptocurrency mixing service.\footnote{Initial transaction hash:  {\tt 96e8c84dfa9dcbe5b161c345877381f2c2\\e83a464c1db1e149c9b0071da9ced8}}
Cryptomixer.io applies a series of transactions similar to a peel chain, a sequence of transactions where a large input is progressively split into smaller outputs across multiple transactions to obscure the origin of the funds and to withdraw money associated to other users of the service. This address served to load the addresses with sufficient funds to participate in the campaign. This was necessary, as some wallets were empty and some amount of funds is of course necessary to transact. These transactions did not include an \opret\ output and thus did not show up in our initial batch of transactions. The attribution of this address to Cryptomixer.io is based on labels obtained from ScoreChain~\cite{scorechain}.

\subsection{Payment Transactions}
Along the \opret\ outputs, the actor also sent small amounts to outputs which in our dataset are all identified as either GRU, SVR, or FSB wallet addresses. The outputs contain fractional amounts of Bitcoin below US\$1. It has a parallel with \textit{dusting attacks}, where tiny amounts of BTC, called \textit{dust}, are sent to trace and analyze transactions, aiming to de-anonymize users. By analyzing transactions that include the dust, attackers can then identify which addresses are likely controlled by the same user.

Inpection of individual transactions reveals an interesting feature, suggesting that a scripted scheme. When inspecting the individual transactions, two things stand out. First, all transactions have a single input address, but multiple output addresses. One transaction even counts one input and 880 outputs of 0.00000547 BTC or US\$0.23 each, with an aggregate total value of US\$1,424.09.\footnote{\tt 2deb61815c8aff5fe89c39bd8ab632b1110f70be3b9fba52b1\\f77d68e3bbc622}

\begin{figure}[t]
    \centering
    %trim=left bottom right top
    \includegraphics[trim=4mm 16mm 24mm 2mm, clip, width=\columnwidth]{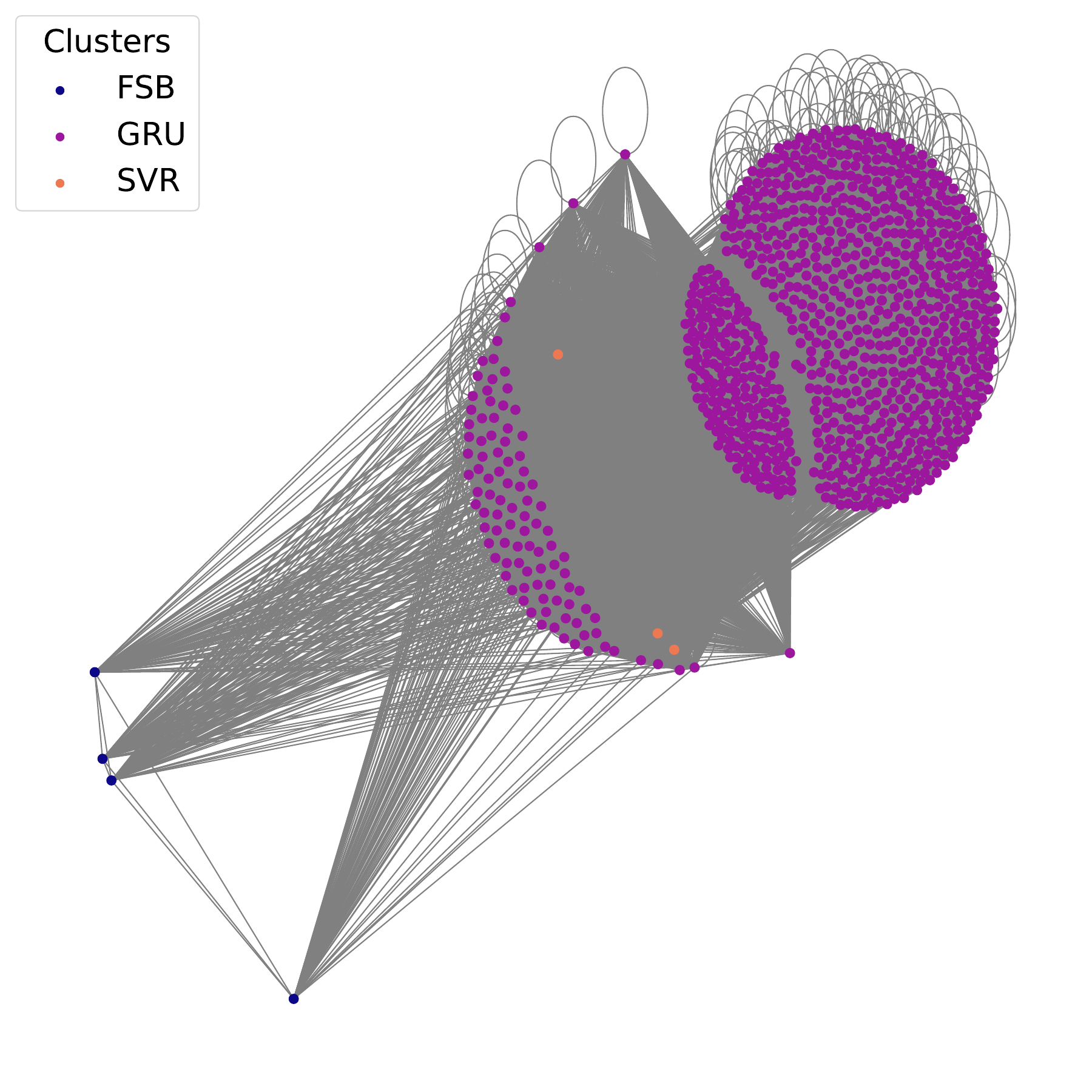}
    \caption{Fig. 2. Force-clustered payment transaction activity.}
    \label{fig:force-cluster}
\end{figure}

\addtocounter{table}{2}
\begin{table*}[t]
\centering
\scriptsize
\caption{Summary of Payment Transactions by Address Attribution}
\begin{tabular}{lrrrrrrrrrr}
\toprule
\textbf{Label} & \textbf{Total} & \textbf{Total Value} & \textbf{Mean Value} & \textbf{Median Value} & \textbf{Min Value} & \textbf{Max Value} & \textbf{Outlier} & \textbf{Outlier Mean} & \textbf{Outlier Min} & \textbf{Outlier Max} \\
 & \textbf{Transactions} & \textbf{(USD)} & \textbf{(USD)} & \textbf{(USD)} & \textbf{(USD)} & \textbf{(USD)} & \textbf{Count} & \textbf{(USD)} & \textbf{(USD)} & \textbf{(USD)} \\
\midrule
\ FSB & 308 & 129.20 & 0.42 & 0.43 & 0.22 & 0.43 & 16 & 0.23 & 0.22 & 0.25 \\
\ GRU & 59,855 & 2,065,728.28 & 34.51 & 0.43 & 0.22 & 8,353.29 & 864 & 1,995.44 & 835.53 & 8,353.29 \\
\ SVR & 1,083 & 336.72 & 0.31 & 0.22 & 0.22 & 0.43 & 0 & 0.00 & 0.00 & 0.00 \\
\bottomrule
\end{tabular}
\label{tab:payments}
\end{table*}

Figure 2 provides a force-clustered overview of the interaction of different GRU, FSB and SVR-labeled wallets during the February 12-18 timeframe. One wallet\footnote{\tt 1594on5HBqWgpxLsvGKdijccdEpxJ5pjZV} is responsible for 100 \opret\ transactions, burning 96,658,067 satoshi, equivalent to 0.966 BTC.

We queried GraphSense~\cite{GraphSense}, a blockchain analysis tool hosted by Iknaio, to discover additional transactions of the addresses involved in the initial transactions between February 12 and 18. Based on that, we identified 11 additional transactions on March 14th with a different \opret\ message, sending funds to the official Bitcoin donation address of the Armed Forces of Ukraine \cite{ukraine}, hosted by Ukrainian cryptocurrency exchange Kuna.io. On March 14th, in 11 transactions with a \textit{Helping Ukraine with money from GRU hackers} \opret\ output, in total US\$975,92 was sent to the official donation address. Of these 11 transactions, which had 637 outputs in total, 11 outputs went to the Ukrainian donation address. The average value of an output was US\$3,22 and the minimum value US\$0.23, again highlighting the circulation of small funds to generate transaction traffic and noise.

\subsection{Ransomware and Breach Activity}
Most addresses in our dataset have never seen activity after the campaign covered in this analysis. However the wallet address \textit{\tt 1EWr1L7BSzFGjk5sZz3zkq5US2x7aiQSJQ}, attributed to the GRU, has been active after 2022. On 24 February 2022, it was observed interacting with a wallet associated with the Conti ransomware group according to labels obtained from Ransomwhe.re \cite{Oosthoek2023}. In one transaction\footnote{\tt f79284691b73c2c667da69a36f648faf4be189a08acadaab05\\4124b9a2fd23cf}, $0.012485$ BTC, worth US\$466.59 at the time, is sent to the group. Notably, the now closed down group has become known for its connection to the FSB \cite{conti-fsb}. In a transaction\footnote{\tt 68a2d5cc511cf08f94b70b774eb11973fd80adf7cae1bdb353\\b5b304d9853792} on May 28th, 2024, the same wallet can be observed sending $0.003302$ BTC, worth US\$229.05 to an address associated with the exploit of the Rain.com cryptocurrency exchange. Finally, on June 26th, 2024 the wallet interacts with Reisbet, a Turkish gambling platform\footnote{\tt 3372f4688cd4bd8207ffceb0a28c54cb7d5b16c1599d000aa4\\3c803ce7a8c741}. The Rain.com and Reisbet labels were provided by ScoreChain \cite{scorechain}.

%https://app.scorechain.com/address/bitcoin/1FXafvUyisk35sAHqmSFmTt6exBXnH3NyE

\begin{figure}[!tpb]\centering
%table included in separate pdf to avoid formatting issues due to Cyrillic text
% trim: left bottom right top
\includegraphics[width=\columnwidth]{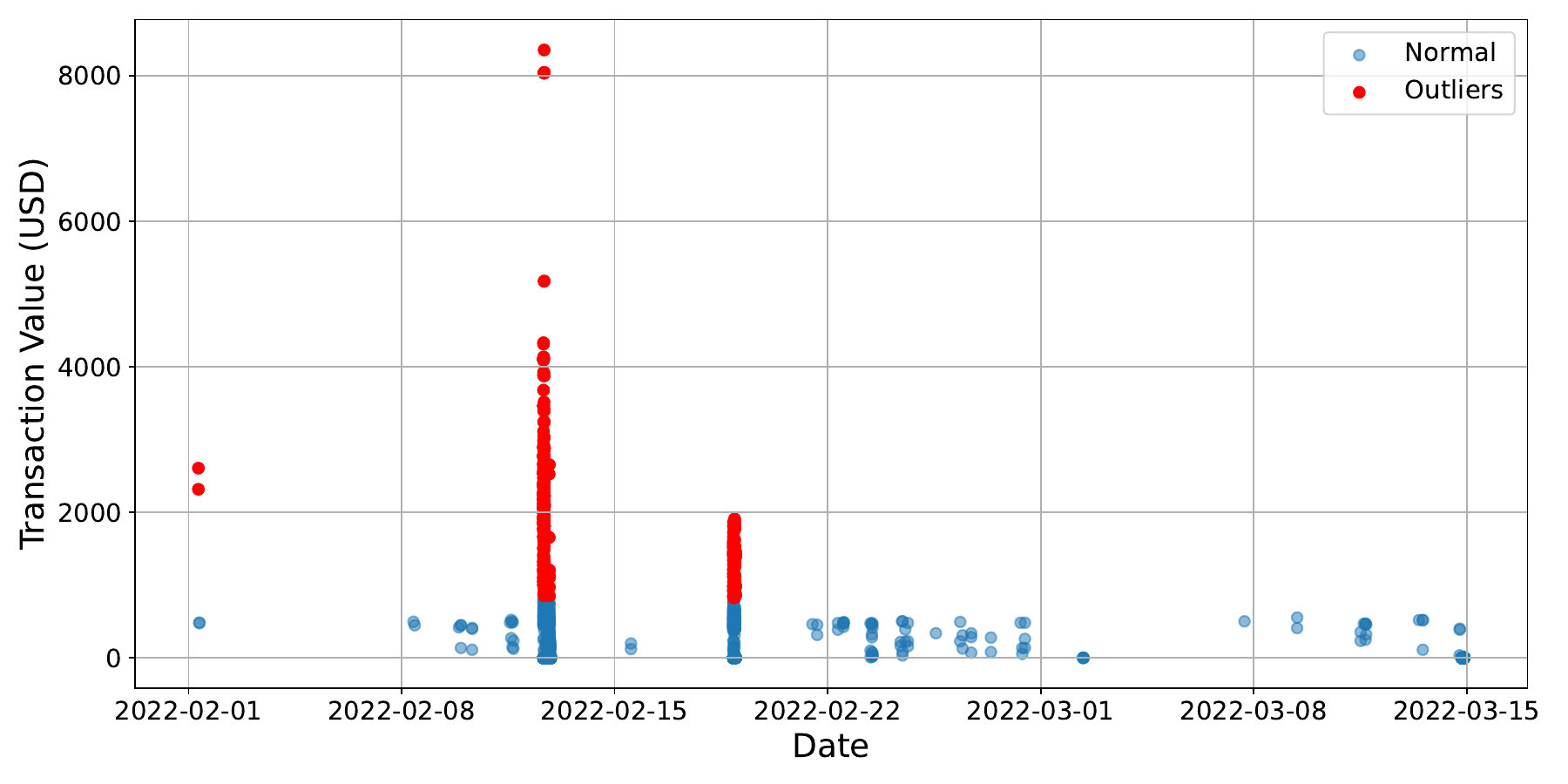}
\caption{Fig. 3. Timeline of Payment Transactions in the OP\_RETURN Campaign, highlighting normal (blue) and outlier (red) values.}
\label{callouts}
\end{figure}
\section{Conclusion}
Within the broader conflict known as the Russo-Ukrainian War, the dates of February 12 and 18, 2022, may appear anomalous, given that Russia’s invasion of eastern Ukraine commenced on February 24, 2022. However, numerous events in the preceding months and days foreshadowed this escalation. As demonstrated in our article, it was in this time frame that the \opret\ actor deliberately generated significant activity on the blockchain by creating a spike and record in both \opret\ outputs and small dust-like transactions. Regarding the \say{6 D's} of cyber warfare (deterrence, deception, disruption, destruction, disinformation, denial), the OP\_RETURN campaign can be classified as an act of denial and destruction. It exemplifies denial by depriving the original owner of access to their financial resources, while the use of the burn opcode in Bitcoin Script effectively destroys those resources.

Building on evidence embedded in the \opret\ binary code suggesting that these are indeed Bitcoin wallets once controlled by Russian intelligence agencies, two primary scenarios emerge regarding the attribution of this campaign. The \opret\ transactions could have been initiated either by a disgruntled insider with direct access or by an outsider who illicitly obtained the private keys. Given that few entities possess the capability, motive, and opportunity to penetrate the security of intelligence agencies such as the FSB, SVR, or GRU, it seems implausible that a low-level attacker — such as a script kiddie -could have acquired the keys and executed this campaign.

Alternative hypotheses regarding the campaign’s origins warrant consideration. One possibility is an inside job, wherein an individual within the GRU, FSB, or SVR — perhaps a disgruntled operator or an IT employee with access to critical systems - misappropriated the Bitcoin. Such an act could stem from motives including financial gain, personal vendettas, or participation in a broader scheme orchestrated by external actors. Another scenario involves a rogue insider collaborating with a third party, providing essential access to systems or expertise in circumventing security protocols, thereby enabling an external hacker to execute the theft.

From a technical perspective, the GRU’s Bitcoin wallet may have been compromised due to a vulnerability — such as a software flaw or an error in cryptographic protocol implementation — allowing an attacker to access and siphon funds. This would indicate a significant oversight on the part of the GRU. Alternatively, a more sophisticated method, such as a man-in-the-middle attack, could have been employed. In this case, the attacker might have intercepted communications during a transaction or wallet transfer process, compromising the GRU’s assets without their immediate awareness.

This is further underscored by the actor's decision to burn over US\$300,000 worth of Bitcoin against the prevailing exchange rate at the time. The choice not to monetize the acquired funds implies the presence of a robust ethical framework. Additionally, it seems highly improbable that Russian actors would voluntarily donate Bitcoin to the Ukrainian cause. The initiator's apparent ability to afford the destruction of over US\$300,000 worth of seized Bitcoin suggests a level of sophistication, simultaneously deterring the original owner from reusing the associated addresses. Consequently, we assess that both the original owner and the actor who appropriated the funds are likely highly skilled actors. However, determining which of the two is more sophisticated, given the potential compromise of the private keys, is a matter for further debate and lies beyond the scope of this paper.

\section*{Acknowledgment}
The authors would like to express their gratitude to Bernhard Haslhofer of Iknaio for providing access to the GraphSense API, which allowed Bitcoin address clustering and also to Scorechain, which facilitated the identification of potential ownership of addresses. This work was supported by the European Commission under the Horizon Europe Programme as part of the project SafeHorizon (Grant Agreement \#101168562). The content of this article does not reflect the official opinion of the European Union. Responsibility for the information and views expressed therein lies entirely with the authors.

\bibliographystyle{IEEEtran}
\bibliography{conference_101719}
\end{document}